\documentclass[aps,prl,showpacs,nobibnotes,twocolumn,
nofootinbib,nobalancelastpage,superscriptaddress]{revtex4}
\usepackage{graphicx}
\usepackage{amsmath}
\usepackage{epstopdf}
\usepackage{latexsym}
\usepackage{dcolumn}
\usepackage{bm}
\input{epsf}
\newcommand{\be}{\begin{equation}}
\newcommand{\ee}{\end{equation}}
\newcommand{\ba}{\begin{eqnarray}}
\newcommand{\ea}{\end{eqnarray}}
\newcommand{\bi}{\begin{itemize}}
\newcommand{\ei}{\end{itemize}}
\newcommand{\ga}{\gtrsim}
\newcommand{\bfi}{\begin{figure}
\epsfxsize=9cm
\epsffile}
\newcommand{\efi}{\end{figure}}

\newcommand{\la}{\lesssim}

\begin{document}
\title{Type Ia supernovae as speed sensors at intermediate
  redshifts}  
\author{Pengjie Zhang}
\affiliation{Shanghai Astronomical Observatory, Chinese Academy of
  Science, 80 Nandan Road, Shanghai, China, 200030}
\affiliation{Joint Institute for Galaxy and Cosmology (JOINGC) of
SHAO and USTC}
\author{Xuelei Chen}
\affiliation{National Astronomical Observatory, Chinese Academy of
  Science, Beijing, China}
\email{pjzhang@shao.ac.cn,xuelei@cosmology.bao.ac.cn}     
\begin{abstract}
Large scale peculiar velocity (LSPV) is  a crucial probe
of dark matter, dark energy and gravity at cosmological scales.
However, its application  is severely limited by
measurement obstacles. We show that fluctuations in type Ia supernovae
(SNe Ia) fluxes induced by LSPV offer a promising approach
to measure  LSPV at intermediate 
  redshifts. In the 3D
Fourier space, gravitational lensing, the dominant systematical error,
is well suppressed, localized and can be further corrected
effectively. Advance in SN observations can further significantly
reduce shot noise induced by SN intrinsic fluctuations, which is the dominant
statistical error. Robust mapping on the motion of the dark universe
through SNe Ia is thus feasible  to $z\sim 0.5$. 
\end{abstract}
\maketitle
\section{Introduction}
Matter distribution of the Universe is being  revealed to great details
by surveys on galaxies, gravitational lensing,
the thermal Sunyaev Zel'dovich (SZ) effect, CMB, X-ray, etc. In
contrast,  measurements on large scale peculiar
velocity (LSPV), or bulk flow, are  still limited. Measurements which
rely on distance indicators to subtract Hubble flow 
 \cite{FP} are limited to  local universe. Those based on anisotropic galaxy
 clustering in redshift space can be extended to cosmological distances
 \cite{Tegmark02}. 
However, this does rely on  modeling of galaxy redshift distortion, whose
accuracy still  requires much improvement in this era of precision cosmology
 \cite{Scoccimarro04}. The
 kinetic SZ (KSZ) effect \cite{Sunyaev80} of clusters is a promising probe
 \cite{KSZ}, although systematic errors do exist 
 \cite{KSZerror}.  Statistics of the diffuse KSZ background, such as
 the power spectrum \cite{KSZps} and cross correlations with other
 tracers of the large scale structure \cite{KSZcross} can be measured
 robustly. However, this background measures the peculiar momentum instead
 and thus probes LSPV only indirectly.

LSPV, as a direct
tracer of gravitational potential at cosmological scales, is powerful
to probe dark matter, dark energy \cite{vcosmology1} and gravity
\cite{vcosmology2}. In fact, it is indispensable  to distinguish between
some scenarios on the dark sectors \cite{Jain07}. We show that it is
promising  to measure LSPV robustly through type Ia
supernovae (SNe Ia) and open a new window into the dark universe.  

Peculiar
velocity of a SN at position $\vec{x}$ shifts the apparent redshift
to $1+z=(1+\bar{z})(1+\vec{v}\cdot\hat{x})$, where $\bar{z}$ is
the real redshift. It also changes the luminosity distance to
$d_L(z)=\bar{d}_L(\bar{z})(1+2\vec{v}\cdot\hat{x})$. Fluctuations
induced in its flux (with respect to the mean flux at $z$, instead of at $\bar{z}$) is then \cite{Hui06}
\be
\label{eqn:v}
\delta_F^v(\vec{x})=Q(z)\vec{v}\cdot\hat{x};\ Q(z)=-2(1-\frac{1+z}{\chi H})\ .
\ee 
Here, $H$ is the Hubble constant at redshift $z$ and $\chi=\int dz/H$
is the comoving distance. 
We have neglected the earth motion, whose effect is straightforward to
take into account.  The signal $\delta_F^v\sim v/(cz)$ is contaminated
by $\delta_F^L$   
induced  by gravitational lensing\footnote{Dust extinction also causes fluctuations in SN fluxes. This kind of
fluctuations has opposite sign to that of gravitational lensing and
thus suppresses contaminations in velocity measurement. For cosmic
gray dust, contaminations can be suppressed by $10$-$50\%$
\cite{Zhang07a}.} and $\delta_F^{\rm
  random}$  induced by SN Ia intrinsic
fluctuations. 
$\delta_F^{\rm  random}$ is spatially
uncorrelated and straightforward to correct. 
$\delta_F^L$ induces systematical errors in velocity measurement,
 however,  at local universe where $z<0.1$, contaminations of
$\delta_F^L$ are negligible ($\langle  
(\delta_F^v)^2\rangle \ga 10^{-4}\gg \langle (\delta_F^L)^2\rangle$).
 These properties  have  enabled success in local  LSPV
 measurements~\cite{SNIameasurement}.  

Gravitational lensing induced systematical error increases with redshift,
while the signal decreases. So eventually the
lensing induced systematical error  will overwhelm the signal at
intermediate redshifts. For example, at $z\sim 0.5$, $\delta_F^v\sim 0.1\%\la
\delta_F^L$.  If not corrected, this will prohibit
the application of SNe Ia as cosmic 
speed censors. However, as the main result of this paper, we will show
that, lensing 
induced systematical error has a intrinsically different pattern to that of
$\delta_F^v$ and  can  be effectively corrected in  the 3D Fourier
space. On the other hand,  shot noise
induced by SNe Ia intrinsic fluctuations, which is the dominant
statistical error,  can be overcome by advance in
observations. This approach does not rely on assumptions on SN Ia absolute
luminosity, other than that its rms fluctuation is small, since it
only explores information imprinted in flux fluctuations. Eventually,
high precision mapping of the motion of the  
dark universe to  $z\sim 0.5$  will be realized through SNe Ia.

For a narrow redshift bin, $\delta_F^v$ and $\delta_F^L$
are virtually uncorrelated, due to the cancellation of positive $v$
and negative $v$ and the lensing weighting function.  We
then have $\langle
\delta_F(\vec{x}_1)\delta_F(\vec{x}_2)\rangle=4w_{\kappa}(\vec{x}_1,\vec{x}_2)+Q^2(z)
\xi_v(\vec{x}_1,\vec{x}_1)$. Here, the direct observable is the total
flux fluctuation (the sum of three)
\be
\delta_F=\delta_F^v+\delta_F^L+\delta_F^{\rm random}\ .
\ee
$w_{\kappa}(\vec{x}_1,\vec{x}_2)\equiv \langle
\kappa(\vec{x}_1)\kappa(\vec{x}_2)\rangle$ and $\xi_v(\vec{x}_1)\equiv \langle
\vec{v}(\vec{x}_1)\cdot \hat{x}_1\vec{v}(\vec{x}_2)\cdot
\hat{x}_2\rangle$ are the correlation functions of $\kappa$ and
$\vec{v}(\vec{x})\cdot\hat{x}$,
respectively. $\kappa$ is
the lensing convergence of a SN Ia at position $\vec{x}$ and
$\delta_F^L=2\kappa$ in the weak lensing regime. $w_{\kappa}(\vec{x}_1,\vec{x}_2)$ is anisotropic with
respect to $\vec{x}_1-\vec{x}_2$ \footnote{During preparation of this
  work, A. Vallinotto et al. \cite{Vallinotto07} and L. Hui et
  al. \cite{Hui07} published works on 3D lensing correlation
  function. Especially in \cite{Hui07} and a pioneer work \cite{Matsubara00} by
  T. Matsubara that \cite{Hui07} pointed out,  the anisotropy in the
  correlation function  $w_{\kappa}(\vec{x}_1,\vec{x}_2)$ is explicitly
calculated. The result presented in this 
  paper is performed in Fourier space, complementary to theirs and
  particularly suitable for LSPV reconstruction that
  this paper focuses on. Recently, Hui et
  al. \cite{Hui07b} 
published their result in Fourier space, which is consistent with ours.}. It
depends  primarily on $\vec{\theta}_{12}$, the angular separation of two lines
of sight (l.o.s.). The 2D
Fourier transform over $\vec{\theta}_{12}$ (under the flat sky
approximation) gives the usual 2D lensing power spectrum
$C_l^{\kappa}(x_1,x_2)$. On the other 
hand,  $w_{\kappa}(\vec{x}_1,\vec{x}_2)$ only weakly depends on $x_1-x_2$, 
the radial separation along the l.o.s..  Namely
$w_{\kappa}$ lacks small scale power along the l.o.s. 
Correspondingly,  the 3D power spectrum $P^{\kappa}_{3D}(\vec{k})$, which is
the 3D Fourier transform of $w_{\kappa}$, is highly anisotropic in 
the 3D  wavevector $\vec{k}$ space. Its power should  concentrate on a 2D
plane of 
$\vec{k}$ perpendicular to the l.o.s., while its power along the
l.o.s. can be significantly suppressed. On the other hand,
$P^v_{3D}(\vec{k})$, 
which is the 3D power spectrum of $\hat{v}\cdot\hat{x}$, has the opposite
behavior.  Its power concentrates along the l.o.s.. This brings hope to measure
$P^v_{3D}(\vec{k})$ along the l.o.s.. 

\bfi{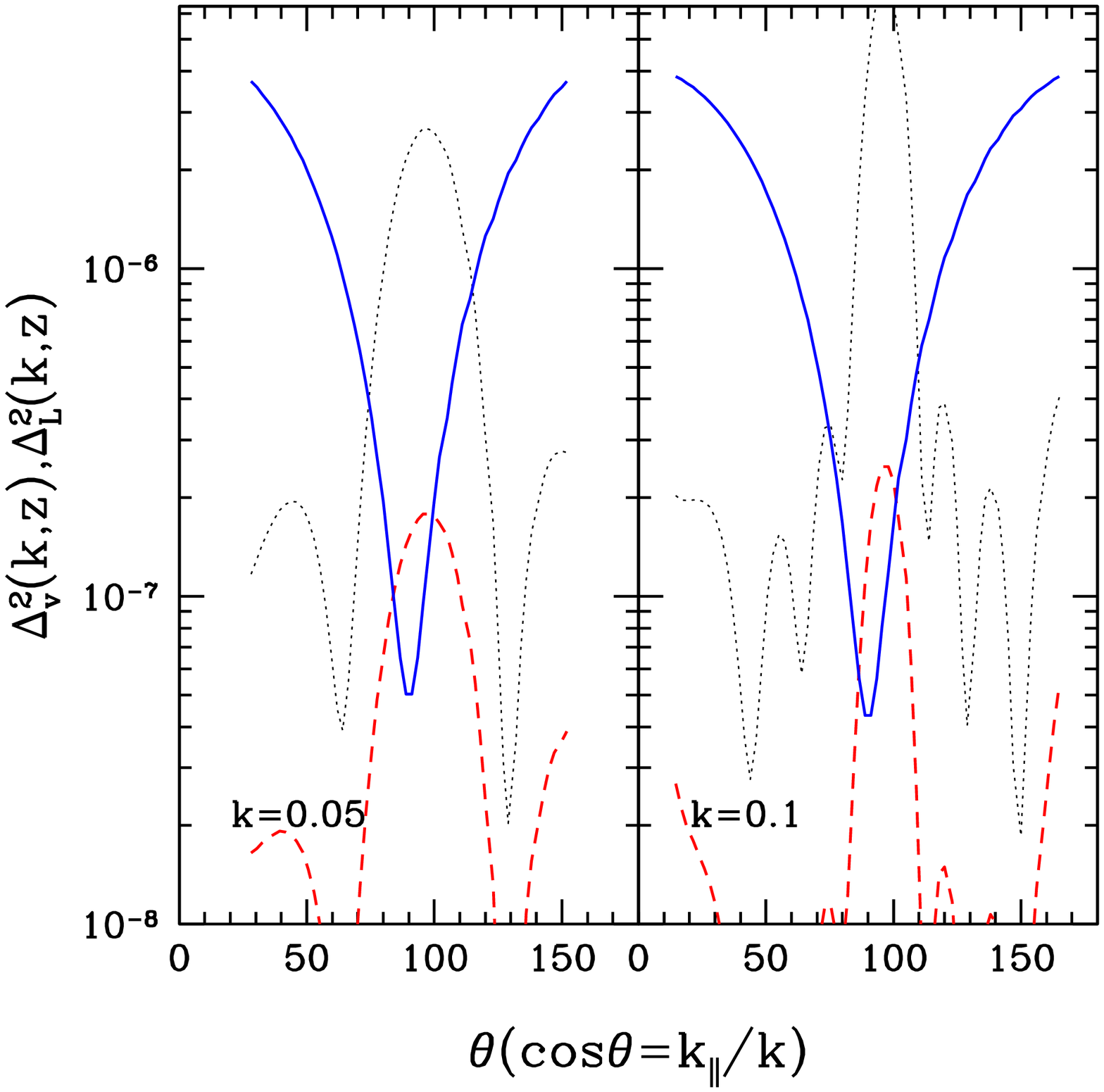}
\caption{The 3D power spectrum variance of SNe Ia flux fluctuations, which is
  the sum of $\Delta^2_v(\vec{k})\equiv
  Q^2(z)k^3P_{3D}^v(\vec{k})/(2\pi^2)$ (solid lines) and $\Delta_L^{2}(\vec{k}) \equiv
 4 k^3P_{3D}^{\kappa}(\vec{k})/(2\pi^2)$ (dot lines), at
  $z\in[0.45,0.55]$.  $k$ is in unit of $h/$Mpc. Gravitational lensing
  lacks small scale power along the l.o.s., which causes the oscillating
  features. The same reason causes significant suppression in
  $\Delta_L^{2}(\vec{k})$  at 
  $k_{\parallel}\gg 2\pi/\Delta x$, where LSPV can be measured accurately in a
  model independent way. When 
  $k_{\parallel}\rightarrow 0$, $\Delta_L^{2}(\vec{k})$ is barely
  suppressed and  $\Delta^2_v(\vec{k})$ can not be measured for these
  modes. $\Delta^2_L$ begins to increase where $k_{\perp}\rightarrow
  0$, caused by its dependence on
  $C^{\kappa}_{l=k_{\perp}x_c}$. 
 Noticing that $\Delta^2_L(\vec{k})$ is slightly
  asymmetric with respect to $\theta\rightarrow \pi-\theta$.  The
  ensemble average of $\Delta_L^2$ can be calculated from  independent
  lensing surveys. After subtraction, gravitational lensing only induces
  statistical error (long dash lines), which is estimated combining
  all $\theta_{\rm max}\times\theta_{\rm max}$ patches in half
  sky. \label{fig:1}} 
\efi

\section{The 3D flux fluctuation power spectrum}
In the appendix, we derive the analytical expression
of the 3D power spectra $P^{\kappa}_{3D}(\vec{k})$ and
$P^v_{3D}(\vec{k})$.  These results apply to survey volume in a narrow
redshift range $[z-\Delta z/2,z+\Delta z/2]$ and sufficiently small
sky area $\theta_{\rm max}\times \theta_{\rm max}$. We require  $\Delta
z\ll z$ and $\theta_{\rm max}/2\ll 90^{\circ}$ and choose the coordinate such
that the center of the survey volume is 
$\vec{x}_c=(0,0,x_c)$. For the small sky coverage adopted, the l.o.s. is close
to  the  direction to the survey center, which we denote with subscript
$\parallel$. The directions perpendicular to the direction to the survey
center are denoted with subscript $\perp$. Here we summarize the results.

The lensing convergence $\kappa$ is determined by the matter over-density
$\delta$ projected along the l.o.s. through $ 
\kappa(\vec{x})=\int \delta(x_L\hat{x})W(x_L,x)dx_L$. The lensing
kernel $W(x_L,x)=3\Omega_0H_0^2x_L(1-x_L/x)/2$ where the lens distance
$x_L<x$ and zero  otherwise.  $w_{\kappa}(\vec{x}_1,\vec{x}_2)$,
$C^{\kappa}_l(x_1,x_2)$, the 3D matter power spectrum $P_m(k,z)$ and the 3D
lensing power spectrum  $P_{3D}^{\kappa}(\vec{k})$ are
related by the following equations, 
\be
\label{eqn:w}
w(\vec{x}_1,\vec{x}_2)=\int \frac{d^2 l}{(2\pi)^2}C^{\kappa}_l(x_1,x_2)
  e^{-i\vec{l}\cdot (\hat{x}_{1,\perp}-\hat{x}_{2,\perp})}\ ,
\ee
\be
\label{eqn:cl}
C^{\kappa}_l(x_1,x_2)=\int
P_m(\frac{l}{x_L};z_L)\frac{W(x_L,x_1)W(x_L,x_2)}{x_L^2}dx_L \ ,
\ee
and
\be
P_{3D}^{\kappa}(\vec{k})= x_c^2\Delta x
\bar{C}_{l=k_{\perp}x_c}\bar{I}(\vec{k})\ .
\ee
Here, $\Delta x$ is the distance interval between $z-\Delta z/2$ and
$z+\Delta z/2$. $\bar{C}_l$ is $C_l(x_1,x_2)$ averaged over the source
distribution $x_{1,2}$. $\bar{I}(\vec{k})$ is  an oscillating
function of $k_{\parallel}$, induced by  the Fourier
transform along the direction of the survey center and reflecting the fact that
$w(\vec{x}_1,\vec{x}_2)$ lacks small scale power along the l.o.s..  When
$k_{\parallel}\gg k_{\perp}\theta_{\rm max}$ and $g\equiv k_{\parallel}\Delta
x/2\gg 1$, $\bar{I}\simeq (\sin g/g)^2\ll 1$. Only when 
$k\rightarrow 0$, $\bar{I}\simeq 1$.

  On the other hand, the 3D
power spectrum of $\delta_F^v$ is
\be
\label{eqn:v3D}
P_{3D}^v(\vec{k},z)\simeq
\frac{\beta^2H^2a^2}{c^2}\frac{P_m(k,z)}{k^4}\left(k_{\parallel}^2+k_{\perp}^2\frac{\langle
    \hat{x}_{\perp}^2\rangle}{2}\right)\ .
\ee
Here $D$ is the linear density growth factor, $\beta\equiv d\ln D/d\ln
a$ and $P_m(k,z)$ is the
matter density power spectrum at redshift $z$. The measured flux fluctuation
power spectrum is the sum of two:  $4P^{\kappa}_{3D}+Q^2(z)P_{3D}^v$.

We show the two 3D power spectra  at
$z=0.5$ in figure \ref{fig:1}. We only show those modes with
$l=k_{\perp}x_c\geq 30$,  where the flat sky approximation holds. 
The power of $P_{3D}^v$ concentrates along  the l.o.s., whereas that
of $P^{\kappa}_{3D}$ can be significantly suppressed. 
Clearly, for modes with sufficiently large $k_{\parallel}$, contribution from
LSPV outweighs that from gravitational lensing by a large factor.
The ensemble average of $P^{\kappa}_{3D}$ can be predicted from
$C^{\kappa}_l$, which will be measured to high precision by upcoming lensing
surveys on the same sky. By subtracting it from the measured flux
power spectrum, the systematic error $P^{\kappa}_{3D}$ is converted
into statistical error and one obtains an unbiased measure of 
$P^v_{3D}$ (Fig. \ref{fig:1}). 

One still needs to overcome SNe Ia intrinsic
  fluctuations, which induce the dominant statistical error,
\ba
\frac{\Delta P_{3D}^v}{ P_{3D}^v}\simeq&
\sqrt{\frac{(2\pi)^2}{k^2\Delta k f_{\Omega}V}}
 \frac{\sigma^2_{\rm intr}}{\bar{n}_{\rm SN} \Delta_v^2
 2\pi^2/k^3}\\
=&0.4 (\frac{k}{0.05})^{3/2} 
 \frac{10^{-3}}{\bar{n}_{\rm SN}} 
(\frac{\sigma_{\rm intr}}{0.1})^2 \frac{3\times
 10^{-6}}{\Delta_v^2}(\frac{10^9}{V} \frac{k}{\Delta kf_{\Omega}})^{1/2}\ .\nonumber 
\ea 
Here, $f_{\Omega}$ is the fractional solid angle of ${\bf k}$ modes
used for the analysis. 
$10\%$ rms  flux fluctuation ($\sigma_{\rm intr}=0.1$) corresponds to
roughly $0.1$ mag rms dispersion in magnitude.  Better calibration of
 SNe Ia and improvement over survey noise are likely able to reduce
 $\sigma_{\rm intr}$ \cite{Wang05}.  A factor of $2$ decrease in $\sigma_{\rm 
 intr}$ would relax the requirement of $\bar{n}_{\rm SN}$ by a factor
 of $4$ or the sky coverage by a factor of $16$. Otherwise, 
successful measurement of LSPV requires (1) large sky coverage
$f_{\rm sky}\sim 1$ and (2) high SN number density $\bar{n}_{\rm
  SN}\sim 10^{-3} (h/$Mpc)$^3$.  There are no fundamental obstacles to
reach  the above goals.  The SNe 
 Ia rate at $z=0.5$ is $\simeq 10^{-4} (h/{\rm  Mpc})^3/{\rm yr}$ 
 \cite{Poznanski07,Kuznetsova07}. So the survey must be performed over
 decades to reach $\bar{n}_{\rm SN}\sim 10^{-3} (h/{\rm Mpc})^3$.
 Drift scan  using  many telescopes of large field of view is  able to
 detect all SNe Ia 
 over a significant fraction of the sky. With advance in technology,
 such survey  can be feasible and economical \cite{ALPACAdiscussion}.
 Follow up measurement on these 
 $\sim 10^6$ spectroscopic redshifts is also feasible, since low
 resolution spectroscopy with $\sigma_v\sim 300$ km/$s$ suffices. Even
 for these surveys, the LSPV measurement  will be still limited to
 directions with sufficiently 
large   $k_{\parallel}$, however, such measurements contain full information on
the 3D power spectrum of $\vec{v}$ (not only $\vec{v}\cdot\hat{x}$), since it
is   isotropic in $\vec{k}$ space.  Furthermore,  since  the power 
of the velocity  power spectrum variance peaks at $k\sim 0.1h/$Mpc for the
standard cosmology, the bulk information of
velocities can be measured. 

Measuring LSPV through SNe Ia requires ambitious improvement over currently
proposed SNe Ia surveys. It thus requires strong justification. Here we
briefly address some impacts of such measurement on precision cosmology. For
cosmology based on general relativity, dark matter and smooth 
dark energy, LSPV measurement helps to break parameter degeneracies. For
example, the strength of weak gravitational lensing increases with dark matter
density $\Omega_m$, because of more lens mass, and $\sigma_8$, because of
stronger density fluctuations. Thus there exists a degeneracy between
$\Omega_m$ and $\sigma_8$. This degeneracy can be broken by the lensing
tomography, through which one can infer the evolution  of the
matter density  field, which  is  more  sensitive to $\Omega_m$  than to
$\sigma_8$.\footnote{In the linear 
  regime, the matter density evolution does not depend on $\sigma_8$. But in
  the nonlinear regime, it does.}  On the other hand, LSPV is very
useful to break this  degeneracy, since the strength of LSPV is
very sensitive to evolution in the matter density field. A simple
justification is as follows. In the linear regime, from the continuity
equation $\dot{\delta}+\nabla\cdot \vec{v}=0$, we have $v(\vec{k},z)\propto
f\delta(\vec{k},z)$ in the Fourier space, where $f\simeq
\Omega_m(z)^{0.56}$. With the measurement 
of $\delta(\vec{k},z)$ (or the power spectrum $P_m(k)$) from weak lensing and
LSPV measurement, one can directly infer $f(z)$ and thus the value of
$\Omega_m$  to break the $\Omega_m$-$\sigma_8$ degeneracy in weak
lensing 
cosmology. LSPV information is also highly complementary to CMB. The
strength of CMB  is sensitive to  the shape and amplitude of  
the initial fluctuations. LSPV measurement directly provides such kind
of 
information and can thus significantly improve the CMB cosmological
constraints. For other applications, refer to \cite{vcosmology1}.

If our universe is indeed fully described by general relativity (GR), dark
matter and smooth dark energy,  cosmological probes other than LSPV
can already put  stringent constraints on dark matter density, dark
energy density and the dark energy equation  of state \cite{DETF}. In this case,
LSPV measurement does not provide fundamentally new information, due
to the fixed (and known) relations between fluctuations in the metric,
density and  LSPV in such cosmology.   However, the dark universe can
be more complicated. If we are 
open to the possibilities that dark  energy may be clustered, with
non-negligible anisotropic stree,  and
gravity at cosmological scales may deviate from GR,   LSPV containes
independent and crucial information to that in flucutations of metric
and density field. LSPV  
measurement can thus provide valuable information, such as the
nature of gravity  at cosmological scales \cite{vcosmology2}. Indeed,
to distinguish  between some dark energy 
models from some modified gravity models, it is indispensable to have LSPV
measurements,  as shown in \cite{Jain07}.

There are several ways of measuring LSPV. LSPV measurement through SNe
Ia is complementary to other  methods. Among them, a promising
one is   the kinetic Sunyaev
Zel'dovich (KSZ)  effect of galaxy clusters.  Ongoing and planned surveys
such as the  south pole  
telescope (SPT)\footnote{http://pole.uchicago.edu/} plus the dark energy
surveys (DES)\footnote{http://www.darkenergysurvey.org/} are able to measure
the LSPV  power spectrum to $10\%$ statistical accuracy at large
scales \cite{KSZ, KSZerror}. Future all sky surveys can further reduce 
the statistical error by a factor of $3$. However, what the cluster
KSZ effect directly 
measures is the total KSZ flux $\propto M_gv$, instead of $v$. Here
$M_g$ is the total gas mass. Some complexities,  such as feedback,
cooling, point source contamination and non-isothermal/non-spherical
gas distribution,  could bias the estimation on $M_g$ \cite{KSZerror}
and thus bias  the LSPV measurement through the
cluster KSZ effect. On the other hand,  LSPV measurement through SNe Ia is
free of such uncertainties, althrough it may suffer some other 
  systematics. For example, if the intrinsic SNe Ia fluctuations are somehow
  correlated with the large scale structure (although quite unlikely), LSPV
  measurements will be biased 
  by such correlation. Nonetheless, SNe Ia
  provide an independent, likely clean and potentially powerful method
  to measure LSPV.   For all these reasons, surveys capable of measuring LSPV
through SNe Ia   would be very profitable scientifically.

\section{Discussions}
Throughout the paper we adopt the flat sky approximation, which breaks where
 $l=k_{\perp}x_c\rightarrow 0$ 
 ($k_{\perp}\rightarrow 0$). However, we will show in a companion
 paper that,  the feature that the power of
$\delta_F^L$  is significantly suppressed along the l.o.s. survives beyond the 
flat sky approximation and so as  the proposed LSPV measurement
 method.

The same technique can be applied to fluctuations in the
fundamental plane relation of galaxies, for which peculiar velocity also
contributes. Dispersion in the 
fundamental plane is larger than that of in SNe Ia. However, higher
galaxy number density may be reached in shorter observation period and
the overall shot noise can be reduced.

 As the summary, this paper shows in a proof of concept study that
contaminations induced by gravitational lensing in SN peculiar
velocity measurements can be efficiently and unbiasedly corrected in 3D Fourier
 space. Combining with advance in SNe Ia surveys, it is feasible to
 measure the large scale peculiar velocity robustly and map the motion
 of the dark universe.

We thank Neal Dalal, Yipeng Jing and Zheng Zheng for useful
discussions. This work is supported by the one-hundred-talents
program of CAS, the NSFC grant  10533010, 10533030, the NSFC
Distinguished Young Scholar grant   No. 10525314, the CAS grant
KJCX3-SYW-N2 and the 973 program grant No. 2007CB815401.

\section{Appendix}
We choose the coordinate such that the center of the survey volume is at
$(0,0,x_c)$.  The 
3D vector pointing to an object is $\vec{x}\equiv x\hat{x}\equiv
(x\hat{x}_{\perp}, x_{\parallel})$. We adopt the 
flat sky approximation. Under this approximation,  the solid angle element
$d\Omega_{\hat{x}}\simeq d^2\hat{x}_{\perp}$ and the volume element $dV\equiv
x^2dxd\Omega_{\hat{x}}\simeq  x^2dxd\hat{x}_{\perp}$. Also,
$x_{\parallel}\simeq x$, accurate to better 
than $2\%$ accuracy, for $20^{\circ}\times  20^{\circ}$ sky.

The 3D lensing power spectrum $P_{3D}^{\kappa}(\vec{k})$ is defined as
\be
\label{eqn:der1}
P_{3D}^{\kappa}(\vec{k})\equiv\int w_{\kappa}(\vec{x}_1,\vec{x}_2)\exp(i\vec{k}\cdot [\vec{x}_1-\vec{x}_2])
\frac{dV_1dV_2}{V} \ .
\ee 
Plug the $w_{\kappa}$-$C_l^{\kappa}$ relation  (Eq. \ref{eqn:w}) into
Eq. \ref{eqn:der1},   integrate over $d^2\hat{x}_{1\perp}$ and
then $d^2l$, we have 
\ba
P_{3D}^{\kappa}(\vec{k})=\int C_{l=k_{\perp}x_1}(x_1,x_2)
e^{i(x_1-x_2)(k_{\parallel}+\vec{k}_{\perp}\cdot\hat{x}_{2,\perp})}
\frac{x_1^2 
dx_1dV_2}{V}\nonumber
\ea
Plug into the Limber integral for $C^{\kappa}_l$ (Eq. \ref{eqn:cl}),
integrate over $dx_{i=1,2}$ and then integrate  over
$d\hat{x}_{2\perp}$, we have 
\ba
\label{eqn:kappa3D} 
P_{3D}^{\kappa}(\vec{k})
&\simeq& x_c^2\Delta x\int
P_m(\frac{k_{\perp}x_c}{x_L},z_L)\frac{dx_L}{x_L^2}\bar{W}^2(x_L)I(\vec{k},x_L)\nonumber\\
&\equiv& x_c^2\Delta x \bar{C}_{l=k_{\perp}x_c}\bar{I}(\vec{k})\ .
\ea
Here we have replaced 
$P_m(\frac{k_{\perp}x_1}{x_L})$ with $P_m(\frac{k_{\perp}x_c}{x_L})$,
since $P_m$ varies slowly with $k$ and $x_1\simeq x_c$ for narrow
redshift bins.  $\bar{W}(x_L)\equiv \int Wx^2dx/\int x^2dx$ is the usual
weighted lensing kernel. The suppression factor 
\ba
\label{eqn:I}
I(\vec{k},x_L)\equiv\bar{W}^{-2}(x_L) \left\langle\left|\frac{\int_{x_l}^{x_u}
W(x_L,x)\exp[ibx]x^2dx}{\int_{x_l}^{x_u} x^2dx}
\right|^2 \right\rangle \nonumber \ .
\ea
Here, $b\equiv k_{\parallel}+\vec{k}_{\perp}\cdot\hat{x}_{\perp}$ and
$\langle\cdots\rangle$ is averaged over the solid angle $d^2\hat{x}_{\perp}$.
The term $bx$ in the 
exponential is $\vec{k}\cdot \vec{x}$ under the flat sky approximation. This implies that, an
orthogonal Fourier mode is along the actual l.o.s. (instead of
along the direction to the survey center).  The other orthogonal
Fourier modes are $\vec{k}_{\perp}$, which is the Fourier transform of
angular direction. $\bar{I}(\vec{k})$ is $I(\vec{k},x_L)$ weighted
  through $x_L\leq x_u$ and its exact definition is given by the last
  relation in Eq. \ref{eqn:kappa3D}. $x_l$ and $x_u$ are the distance at
  $\bar{z}-\Delta z/2$ and $\bar{z}+\Delta z/2$, respectively.

When $x_L\la x_l$,
$I(\vec{k},x_L)$ varies slowly with respect to $x_L$, because $W(x_L,x)$ in the bracket varies slowly with respect to $x\in
[x_l,x_u]$ and cancels the one in $\bar{W}$. When $x_L\rightarrow
x_u$, $I$ increases rapidly and approaches $1$.  Since most lensing 
  contributions come  from  $x_L<x_l$ and the lensing
  weighting ($W^2I$) peaks at $x_L\simeq x_u/2$, a
  convenient approximation is that  $\bar{I}(\vec{k})\simeq
  I(\vec{k},x_L=x_u/2)$. Numerical evaluation of
  Eq. \ref{eqn:kappa3D} shows that  this approximation
  degrades for smaller $k$, however, even for
  $\vec{k}=(0,0,0.05)h/$Mpc, it still works to  $10\%$ accuracy.  We
  will adopt this approximation, for its simplicity and clear physical
  meaning.  

At linear scale, peculiar velocities are irrotational. The mass
conservation equation  $\dot{\delta}+\nabla\cdot
\vec{v}/a=0$ then tells that
$\vec{v}(\vec{k})=i\delta(\vec{k})\vec{k}/k^2$. The 3D (l.o.s.)
velocity power spectrum is  
\ba
P_{3D}^v(\vec{k}^{'})&\equiv& \int \langle \vec{v}(\vec{x}_1)\cdot \hat{x}_1\vec{v}(\vec{x}_2)\cdot \hat{x}_2\rangle
e^{-i\vec{k}^{'}\cdot(\vec{x}_1-\vec{x}_2)} \frac{dV_1dV_2}{V}\nonumber\\
&\simeq &\int 
\frac{d^3k}{(2\pi)^3}\frac{P_m(k,z)}{k^4}e^{i(\vec{k}-\vec{k}^{'})(\vec{x}_1-\vec{x}_2)}
\frac{dV_1dV_2}{V}  \nonumber\\
&& \left(k_{\parallel}^2+(k_{\perp}\cdot
\hat{x}_{1,\perp})(k_{\perp}\cdot
\hat{x}_{2,\perp})+2k_{\parallel}k_{\perp}\cdot
\hat{x}_{\perp})\right)
 \nonumber\\
&=&\frac{P_m(k,z)}{k^4}k_{\parallel}^2+  \int 
\frac{d^3k}{(2\pi)^3}\frac{P_m(k)}{k^4}\frac{dV_1dV_2}{V}\nonumber\\
&&
\left((k_{\perp}\cdot \hat{x}_{\perp})^2-(k_{\perp}\cdot 
\frac{\Delta \hat{x}_{\perp}}{2})^2\right)
e^{i(\vec{k}-\vec{k}^{'})(\vec{x}_1-\vec{x}_2)}\nonumber\ .
\ea
Here, $\hat{x}_{\perp}\equiv
(\hat{x}_{1\perp}+\hat{x}_{2\perp})/2$, $\Delta
\hat{x}_{\perp}\equiv \hat{x}_{1\perp}-\hat{x}_{2\perp}$ and
$\vec{x}_{1,\perp}-\vec{x}_{2,\perp}\simeq x_c\Delta
\hat{x}_{\perp}$.  Finally we have
\ba
\label{eqn:fullv}
P_{3D}^v(\vec{k}^{'})&=&\frac{\beta^2H^2a^2}{c^2}\frac{P_m(k,z)}{k^4}k_{\parallel}^2+\frac{\beta^2H^2a^2}{c^2}\\
&\times&\left(\frac{P_m(k,z) k_{\perp}^2\langle
    \hat{x}_{\perp}^2\rangle}{2k^4}+\left[\frac{P_m(k,z)k_ik_j}{k^44x_c^2}\right]_{,ij}\right)\nonumber\ .
\ea


Nonlinearity at small scales could modify  the above result.  For example,
shell crossing generates vorticity (rotational velocity component), whose
impact  on  $P_{3D}^v(\vec{k})$  could be non-negligible where
$\vec{k}_{\perp}\neq 0$. However, such possible modifications do not invalidate
the proposed LSPV measurement approach. Since the 
main purpose of this paper is  to 
demonstrate the feasibility of measuring LSPV through SNe Ia, instead of
developing an advanced LSPV model, Eq. \ref{eqn:fullv} suffices.  Furthermore,
due to the extra factor 
$(kx_c)^{-2}$, the last term in Eq. \ref{eqn:fullv} is much smaller than the
second term for relevant $k$ and will be neglected elsewhere in this paper.

\newcommand\AAP[3]{A\&A {\bf #1}, #2~ (#3)}
\newcommand\AJ[3]{AJ.{\bf #1}, #2~(#3)}
\newcommand\APJ[3]{ApJ {\bf #1}, #2~ (#3)}
\newcommand\APJL[3]{ApJL {\bf #1}, L#2~(#3)}
\newcommand\APJS[3]{ApJS {\bf #1}, #2~(#3)}
\newcommand\MNRAS[3]{MNRAS {\bf #1}, #2~(#3)}
\newcommand\PRL[3]{PRL {\bf #1}, #2~(#3)}
\newcommand\PRD[3]{PRD {\bf #1}, #2~(#3)}


\begin{thebibliography}{10}
\bibitem{ALPACAdiscussion} We only briefly mention one design which may fulfill
  the survey requirement.  ALPACA \cite{ALPACA}, a proposed $8$ meter
  rotating liquid mercury mirror telescope for SNe Ia search, has
  $\sim 10$ deg$^2$ field of view. $\sim 100$ of
  such telescopes  can be constructed with reasonably low cost and
  detect all SNe Ia over the whole sky.
\end{thebibliography}
\end{document}